\documentclass{pasj00}
\draft

\usepackage{color}

\begin{document}
\SetRunningHead{K. Ishikawa et al.}{X-ray observation of Mars with Suzaku}
\Received{2011/05/20}
\Accepted{2011/08/29}
\Published{}

\title{X-ray Observation of Mars with Suzaku \\at Solar Minimum}
\author{Kumi \textsc{Ishikawa}, Yuichiro \textsc{Ezoe}, Takaya \textsc{Ohashi}, }
\affil{Department of Physics, Tokyo Metropolitan University, 1-1 Minami-Osawa, \\ Hachioji, Tokyo, 192-0397}
\email{kumi@phys.se.tmu.ac.jp, ezoe@tmu.ac.jp, ohashi@tmu.ac.jp}

\author{Naoki \textsc{Terada},}
\affil{Department of Geophysics, Tohoku University, 6-3 Aoba, Aramakiaza, Aoba, Sendai, Miyagi, 980-8578}
\email{teradan@stpp.gp.tohoku.ac.jp}
\and
\author{Yoshifumi \textsc{Futaana}}
\affil{Swedish Institute of Space Physics, Box 812, SE-98128 Kiruna, Sweden}
\email{futaana@irf.se}

\KeyWords{planets and satellites: individual (Mars) --- X-rays: individual (Mars)} 

\maketitle

\begin{abstract}

Mars was observed in X-rays during April 3$-$5 2008 for 82 ksec 
with the Japanese Suzaku observatory. Mars has been known to emit X-rays via the 
scattering of solar X-rays and via the charge exchange between neutral atoms in the 
exosphere and solar wind ions. Past theoretical studies suggest that the exospheric 
neutral density may vary by a factor of up to 10 over the solar cycle. 
To investigate a potential change of the exospheric charge exchange emission,
Mars was observed with Suzaku at solar minimum.
Significant signals were not detected at the position of Mars in the energy band of
0.2$-$5 keV. A 2$\sigma$ upper limit of the O\emissiontype{VII} line flux in 
0.5$-$0.65 keV was 4.3$\times$10$^{-5}$ ph cm$^{-2}$ s$^{-1}$. Comparing 
this upper limit to the past Chandra and XMM-Newton observations conducted
near solar maximum, it was found that the exospheric density at solar minimum 
does not exceed that near solar maximum by more than 6$\sim$70 times.

\end{abstract}

\section{Introduction}

Martian X-ray emission was discovered in 2001 July with the ACIS-I 
X-ray CCD detector onboard the Chandra observatory \citep{den02a}. 
Thanks to the superb X-ray angular resolution 
of 0.5 arcsec, the X-ray emission was resolved into two components. The first one is
seen at the position of the Martian atmosphere. 
The observed X-ray spectrum was dominated by a single narrow emission 
line, which was identified as the 0.53 keV oxygen K$_\alpha$ fluorescence line; 
the total X-ray luminosity due to fluorescence was $\sim$4 MW.
These spatial and spectral characteristics strongly suggest scattering of 
solar X-rays in the upper Martian atmosphere, mainly by fluoresicence. 
This type of X-ray emission has been observed in many other planets like Venus, Earth, 
Jupiter and Saturn (\cite{den02b,gra68,gla02,nes04}). 

The second component was seen as a faint X-ray halo around Mars that 
could be traced out to three Mars radii. Within the very limited statistical quality, 
the overall shape of its X-ray spectrum could be characterized by thermal bremsstrahlung 
emission with a temperature of 0.2 keV. From the analogy with the X-ray emission from 
comets (\cite{den97,weg98}), charge exchange emission between high energy solar 
wind ions and exospheric neutrals was suggested.

The subsequent observation of Mars was conducted on 2003 November 
19$-$21 with XMM-Newton \citep{den06}. From the high energy resolution 
spectroscopy with the Reflection Grating Spectrometer, the presence of $\sim$12 
emission lines was revealed in the Martian halo. The emission lines were considered 
to be due to de-excitation of highly ionized C, N, O, and Ne atoms. 
The He$-$like O\emissiontype{VII} triplet was found to be dominated by the spin-forbidden 
magnetic dipole transition (2$^3$S$_1$ $\rightarrow$ 1$^1$S$_0$). Since the forbidden 
line is weak compared to the resonance line in astrophysical plasmas 
\citep{kra04}, this is a supporting evidence that the Martian halo X-ray emission originates 
from the charge exchange reaction. 

Theoretically, an X-ray flux of the exospheric solar wind X-ray emission is 
proportional to the incident solar wind flux and the neutral column density of the 
collisionally thin regions in the exosphere, that is mainly composed of hydrogen, 
in the line of sight \citep{cra97}.
Therefore, the X-ray observation of the Martian halo emission can be a probe to investigate the 
spatial distribution of hydrogen atoms in the exosphere \citep{hol01} and must be useful to study
the exosphere and also an atmospheric escape from Mars. 
A series of exospheric density models based on measurements with Mariner 6, 7, 9, 
Viking 1, EUVE, Phobos 2, and Mars Express have been constructed 
\citep{and74,kra96,gal06,chau08}. 
Some models \citep{kra96,kra02} of the exospheric density and temperature based on 
the measurements by Mariner 6, 7 and Viking 1 imply a high dependence on the solar cycle. 
They suggested that hydrogen exospheric density at solar minimum can be higher than 
solar maximum due to weaker heating by solar ultra-violet lights and X-rays. 
On the other hand, there are some opposite results measured by Mars Express at low 
solar activity. They indicated no trends to higher exospheric density for low solar activity, 
comparing to the Mariner 6 and 7 data at solar maximum \citep{gal06,chau08}.
Then, the density of the spatially extended thin atmosphere is not well understood because 
of the observational difficulty of the exosphere.  As a result, a possible variation depending 
on the solar cycle is unclear. 

From this point of view, in this paper, we observed Mars in X-rays for the 
first time at solar minimum using the Japanese Suzaku observatory. This data set provides
us with a good opportunity to compare the Martian halo emission at solar minimum with those
observed with Chandra and XMM-Newton near solar maximum.

\section{Observation}
\label{sec:obs}

Suzaku \citep{mit07} observed Mars from 2008 April 3, 08:11 to April 5, 13:00 
UT with the X-ray Imaging Spectrometer (XIS; \cite{koy07}). The XIS consists of three 
front-illuminated (FI) CCDs (XIS0, 2, 3) and one 
back-illuminated (BI) CCD (XIS1). These CCDs cover the energy range of 0.2--12 keV. 
Due to the low-Earth orbit of Suzaku and large effective area, the XIS has one of the lowest 
particle backgrounds among all X-ray CCDs in currently available X-ray observatories. 
Therefore, Suzaku is favorable for the spatially extended emission of Mars, i.e., the Martian
halo emission. In this paper, we used the XIS 0, 1, and 3 data. The XIS2 data was unusable
at the time of this observation due to, most probably, micro-meteoroid impacts.\footnote{See \S 7.12 in the Suzaku technical description document \\(http://www.astro.isas.ac.jp/suzaku/doc/suzaku\_td/node10.html)}

During the Suzaku observation, heliocentric and geocentric distances of Mars were 1.7 and 
1.4 AU, respectively. The average phase angle (Sun-Mars-Earth) and the elongation angle 
(Sun-Earth-Mars) was 37.0 degree and 87.2 degree, respectively. 
Mars had an apparent diameter of 6.8 arcsec and a visual magnitude of m$_{\rm V}$ was 0.86. 
An optical loading to the XIS was thought to be negligible in this observations,
because no optical loading effect was seen when Suzaku observed an optically brighter 
object, i.e., Jupiter \citep{ezo10}. Although the visual surface brightness of Mars was brighter 
than Jupiter during the past Suzaku observation (Jupiter and Mars were 5.5 and 4.7 mag 
per square arcsec during each Suzaku observation, respectively), the effective surface 
brightness of Mars when observed with Suzaku should be lower because the apparent 
optical size of Jupiter is larger than that of Mars (the apparent diameter of Jupiter
was 39 arcsec) and the images are smeared with the Suzaku's angular resolution of 120 arcsec. 

Since Mars moved 27 arcmin per day during the observation, the spacecraft was repointed 27 
times in 3 days to keep Mars within $\sim$1.5 arcmin from the center of the XIS field-of-view (FOV)
in order to avoid the vignetting effect, as summarized in table \ref{tbl:obs}. 
Considering the small offset 
angle from the FOV center, the vignetting effect is almost negligible (See figure 11 in \cite{ser07}). 
The XIS CCDs were operated in the normal mode. The net exposure time after the standard data 
screening was 82 ksec. 
We analyzed the screened data with the process of version 2.2.7.18 provided by the Suzaku 
processing facility, using the HEAsoft version 6.4 package\footnote{http://heasarc.nasa.gov/lheasoft/}.

Figure \ref{fig:mars-all} shows an XIS1 image in the 0.2--1 keV band. 
We adopted the XIS BI for the energy range of 0.2--1 keV and the FI for 1--5 keV, 
since the BI is more sensitive to $<1$ keV while XIS FI is to $>1$ keV.
The first pointing was toward (RA, Dec)$=$(102$^{\circ}$.886, 25$^{\circ}$.171, J2000), ($l$, $b$)
$=$(190$^{\circ}$.245, 11$^{\circ}$.205), while the last one was toward (RA, Dec)$=$(103$^{\circ}$.964, 
25$^{\circ}$48), ($l$, $b$)$=$(190$^{\circ}$.776, 12$^{\circ}$.040). No clear signals were detected 
at the expected positions of Mars (green circles) except for several point like sources in the hard
energy band.

\section{Image}
\label{sec:img}

\begin{figure}[p]
\begin{center}
\FigureFile(100mm,80mm){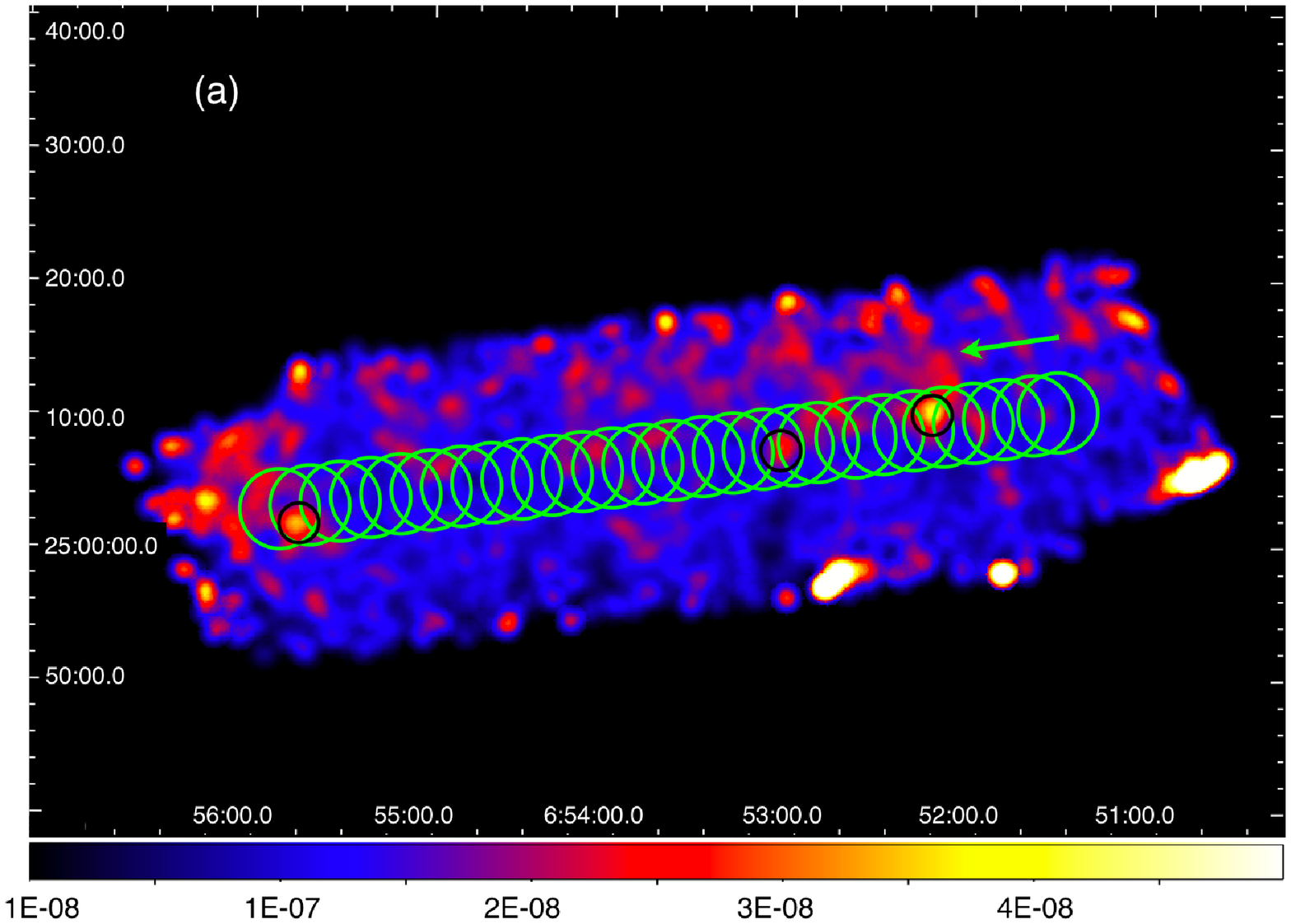}
\bigskip

\FigureFile(100mm,80mm){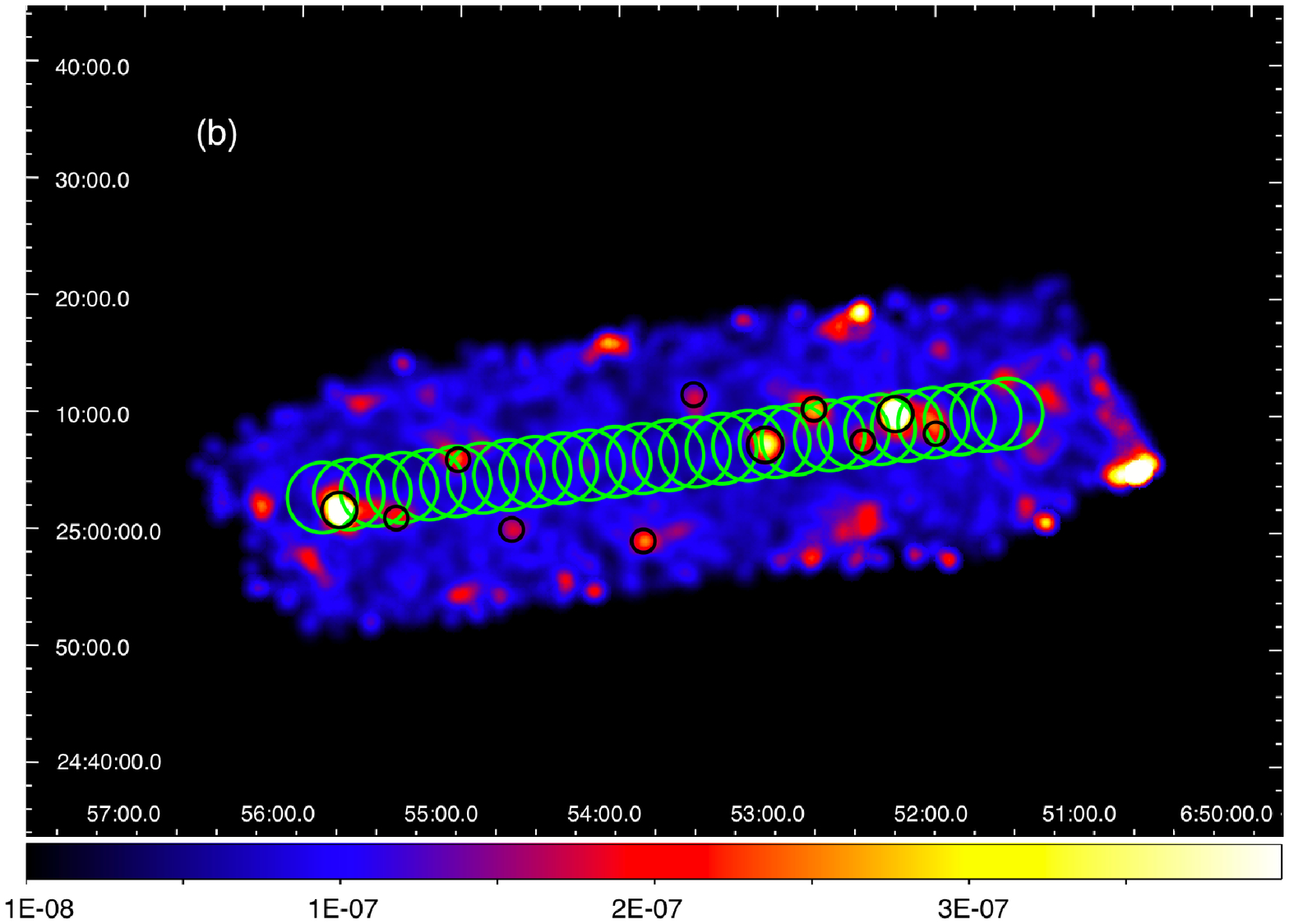}

\caption{
An XIS mosaic image in 0.2--1 keV (panel a, BI) and 1--5 keV (panel b, FI)
displayed on the J2000.0 coordinate.
Images are corrected by exposure time and a count unit is counts s$^{-1}$ binned 
pixel$^{-1}$. For clarity, the image is binned by a factor of 8 (8 arcsec) 
and smoothed by a Gaussian of $\sigma$ = 5 pixels (5.3 arcsec).
An arrow indicates the moving direction of Mars. 
A green circle displays the positions of Mars. 
Its radius is 3 arcmin, corresponding the sum of HPD (Half 
Power Diameter; 2 arcmin) of Suzaku and the moving step 
for the attitude shift.
Black circles represent possible point sources that are excluded for further analysis. 
}
\label{fig:mars-all}
\end{center}
\end{figure}

In order to quantify the Martian X-ray emission in the X-ray images, 
we have to carefully exclude possible point sources in the trail of Mars 
and correct the data for the orbital motion of Mars.
As a first step, we searched the image for bright sources (e.g. cluster of galaxies, 
supernova remnants and stars) using the bright X-ray source catalog compiled by 
NASA\footnote{http://heasarc.nasa.gov/cgi-bin/Tools/high\_energy\_source/high\_energy\_source.pl}.
We found no point sources within the XIS FOVs in this catalog. 
Then, we identified possible faint point sources using the wavelet function program {\tt wavdetect} 
in the CIAO package\footnote{http://cxc.harvard.edu/ciao/index.html}.
We created two images for 0.2--1 keV and 1--5 keV with different bin sizes (4.4 and 8.8 arcsec) 
to detect soft and hard X-ray sources and to take into account different point spread functions 
within the FOV. We ran the program by setting the significant threshold above 1$\times$10$^{-6}$, 
which corresponds to one spurious source in each XIS FOV (1024$\times$1024 pixels).
We detected eleven point source candidates as shown in figure \ref{fig:mars-all} (black circles).
From the individual images in the 27 pointings, we confirmed that all these sources
are bright in X-rays in relevant to the Martian orbit and the position of any 
point source candidate is fixed on the sky.
Hence, we rejected a possibility that the point source candidates include short time 
variation(s) of the Martian X-ray emission. 
Their X-ray fluxes are 0.6$\sim$1$\times$10$^{-13}$ erg cm$^{-2}$ s$^{-1}$ in 
0.2--1.0 keV and 0.1$\sim$2$\times$10$^{-13}$ erg cm$^{-2}$ s$^{-1}$ in 1--5 keV. 
Here we converted an XIS count rate into flux, assuming a power-law spectrum with a photon index of 
1.4 and an average absorption column toward this field\footnote{http://heasarc.gsfc.nasa.gov/cgi-bin/Tools/w3nh/w3nh.pl}, 
which represents a typical spectrum of the background active galactic nucleus.
For comparison, we estimated the number of CXB (Cosmic X-ray Background) sources within the field 
of view ($\sim0.37$ deg$^2$) based on \cite{gia01}. The expected numbers became 9 and 90 if we 
assume the flux limits of $1\times10^{-13}$ and $1\times10^{-14}$ erg cm$^{-2}$ s$^{-1}$ in 2--7 keV, 
respectively, while the observed numbers were 3 and 11. Thus, the source number seems inconsistent with 
the CXB estimation. This is more evident for faint sources, which may suggest that a part of the faint 
sources below $1\times10^{-13}$ erg cm$^{-2}$ s$^{-1}$ in 2--7 keV are not point sources but 
fluctuations of the instrument background.
Since we do not know which point sources are real, we decided to exclude all the X-ray sources 
in the following analysis, in order to avoid potential contamination. We excluded circular 
regions around individual sources with a diameter of 2 arcmin except for three bright sources, 
for which a 3 arcmin diameter region was utilized.

\begin{figure}[p]
\begin{center}

\FigureFile(100mm,80mm){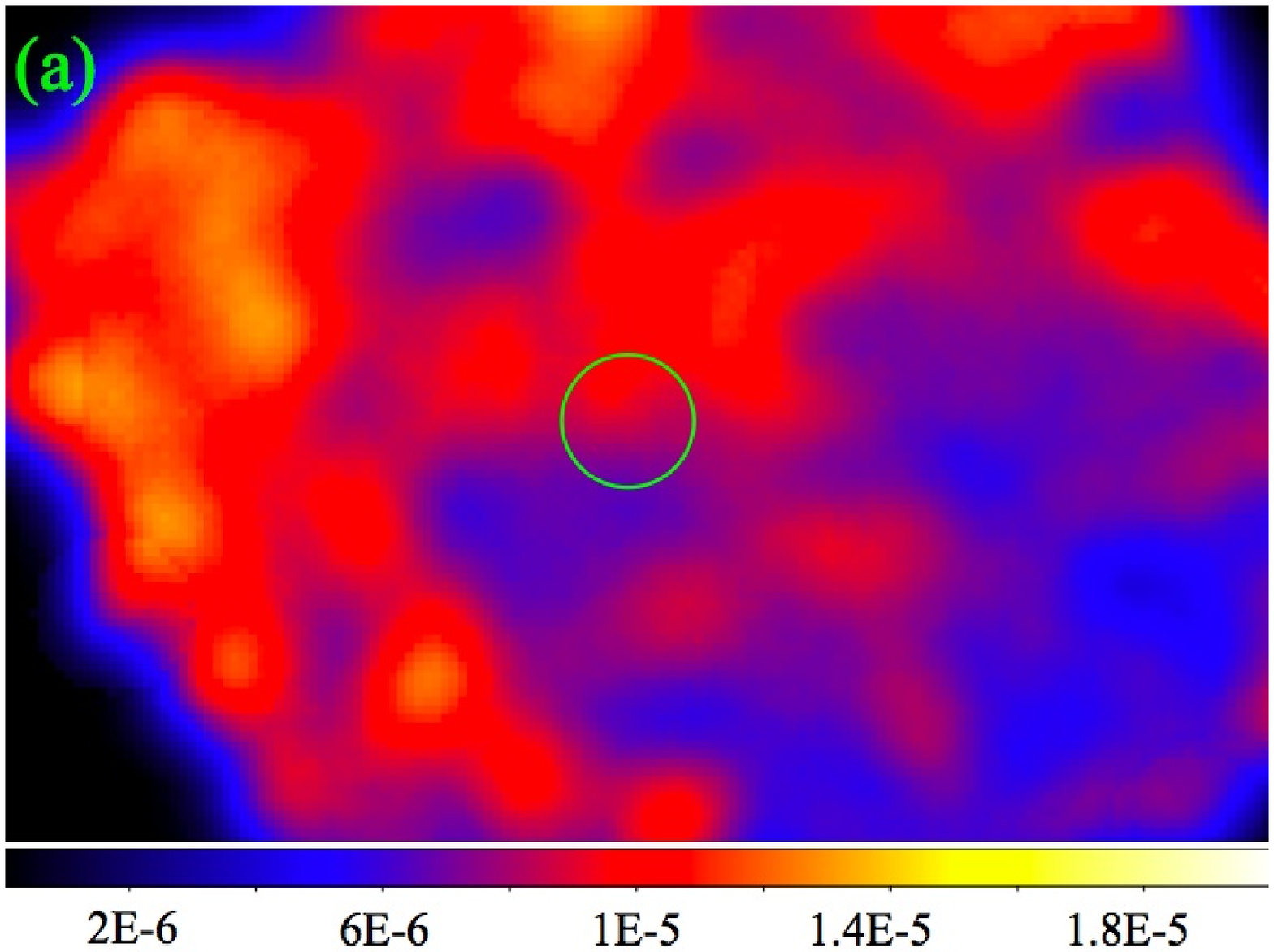}
\bigskip

\FigureFile(100mm,80mm){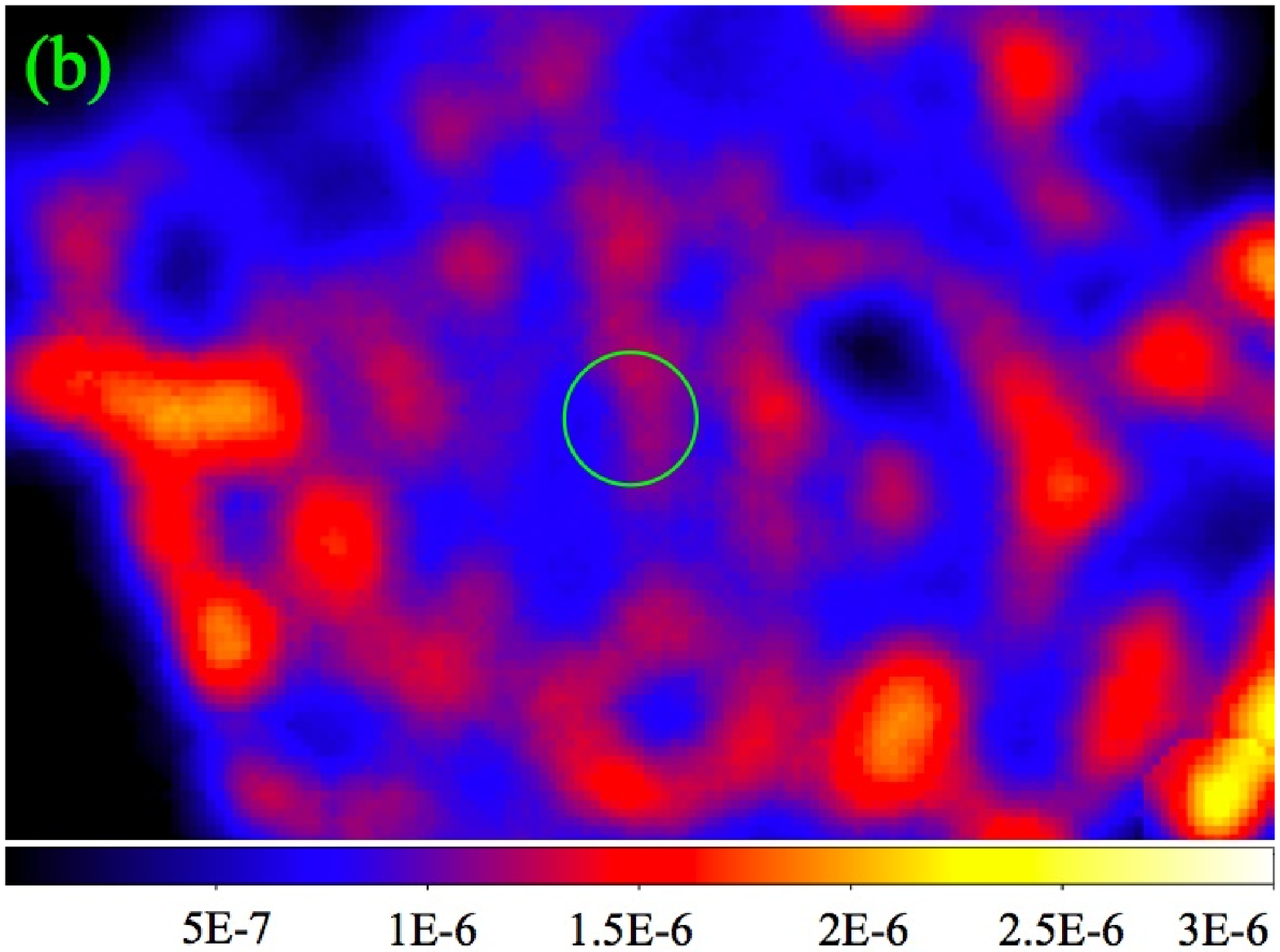}

\end{center}
\caption{XIS images after correction for the orbital motion of Suzaku and Mars's 
ephemeris (http://ssd.jpl.nasa.gov/horizons.cgi) in 0.2--1 keV (panel a) and 
in 0.5--0.65 keV (panel b). Images are binned and smoothed in the same way as figure 
\ref{fig:mars-all}. Green circles indicate the expected position of Mars. Its diameter 
is 2 arcmin considering the HPD and pointing uncertainty of Suzaku ($\sim$20 arcsec 
in radius) \citep{uch08}.}
\label{fig:xis1-dynsky}
\end{figure}

After removing the potential point sources, 
we checked each image and found that no significant
X-ray emission is seen along the orbital path of Mars.
To confirm this inspection, we corrected the images (figure \ref{fig:mars-all})
for the orbital motions of Mars and Suzaku in the same way as \citet{ezo10}. To take
into account the excluded regions, we created exposure maps using {\tt xisexpmapgen}
and excluded the same regions from the exposure maps. Then, we transformed the exposure 
map considering Martian orbit and divided the orbital motion corrected image by the
corrected map. The images corrected for the orbital motions are shown in figure 
\ref{fig:xis1-dynsky}. We chose two energy bands (0.2--1.0 keV and 0.5--0.65 keV),
i.e., a broad charge exchange band and a narrower region, centered on the 
O\emissiontype{VII} line.
However, no significant emission was seen at the position of Mars (green circle) 
in both energy bands. 

\section{Spectrum}
\label{sec:spec}

To place a quantitative upper limit on the X-ray flux of Mars, we proceeded to a spectral analysis.
In the same way as the imaging analysis (\S \ref{sec:img}), we excluded the point source candidates 
from each event file for the 27 pointings without the correction for the orbital motions. 
We then extracted events from a circle region with a radius of 3 arcmin centered on Mars. This radius 
covers the orbital motion of Mars during each pointing ($\sim3$ arcmin) and considers the HPD.
Background events were extracted from an annulus around Mars with a radius of 3 arcmin to 6 arcmin 
after subtracting the point sources.

\begin{figure}[p]
\begin{center}
\FigureFile(100mm,80mm){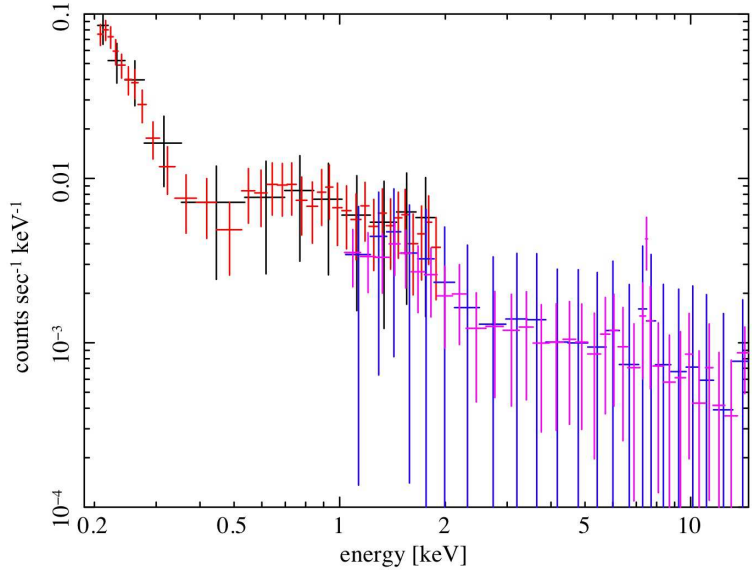}
\bigskip

\end{center}
\caption{XIS spectra of Mars compared to the surrounding background. 
The BI CCD spectra are shown in black (Mars) and red (background),
while the FI spectra are shown in blue (Mars) and magenta (background). 
An error is 1$\sigma$ significance. 
}
\label{fig:spec}
\end{figure}

Figure \ref{fig:spec} shows the obtained X-ray spectra of Mars and the background. 
We can confirm that no significant emission is detected in 0.2--10 keV. A rise around
7--8 keV seen in both Mars and background spectra is most probably an instrumental
Ni K$_\alpha$ line\footnote{http://heasarc.nasa.gov/docs/astroe/prop\_tools/suzaku\_td/node10.html}.
The 2$\sigma$ upper limit of the Martian 
X-ray emission observed near Earth in the O\emissiontype{VII} band (0.5--0.65 keV) was 1.7$
\times10^{-3}$ [cts s$^{-1}$], corresponding to the photon flux of 4.3$\times10^{-5}$ [photon cm
$^{-2}$ s$^{-1}$]. The upper limits in three energy bands are summarized in table \ref{tbl:up}.

\section{Discussion}
\label{sec:discussion}

Below we would like to discuss Martian X-ray emission in the context of the 
scattering of solar X-rays and the exospheric solar wind charge exchange.
In order to comprehensively understand the characteristics of Martian X-ray emission, 
we compared our result with the past Chandra observation at solar maximum \citep{den02a} 
and the XMM-Newton observation at the intermediate state \citep{den06}. 
We also utilized the solar wind data taken with the ACE satellite and the solar X-ray flux 
measured with GOES-12.

\subsection{Scattering of solar X-rays}
\label{sec:sec:scattering}

\begin{figure}[p]
\begin{center}
\FigureFile(100mm,80mm){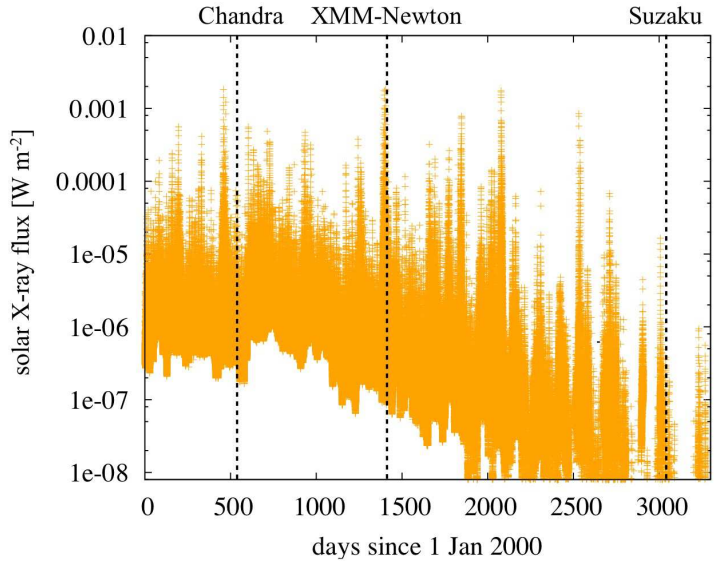}
\end{center}
\caption{The solar X-ray flux from 2000 January 1 to 2008 December 31 measured with GOES-12 1.0-8.0\AA.}
\label{fig:solar-xray}
\end{figure}

\begin{figure}[p]
\begin{center}
\FigureFile(120mm,100mm){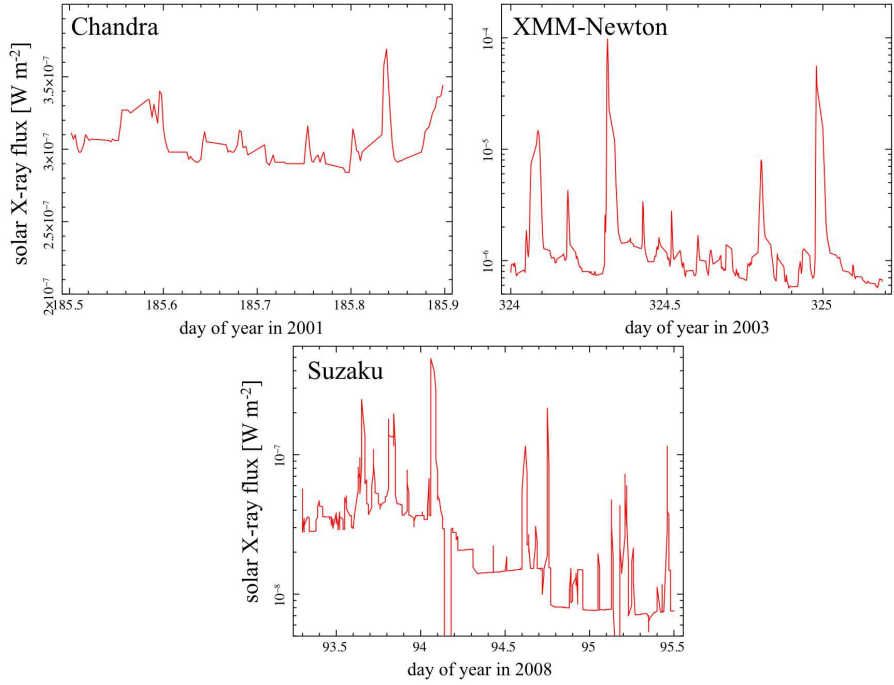}
\end{center}
\caption{The same as figure \ref{fig:solar-xray} but close-up views 
for the Chandra (top left), XMM-Newton (top right), and Suzaku (bottom) observations.
The vertical scale is linear for the Chandra panel but logarithmic for the others.}
\label{fig:solar-xray-CNO}
\end{figure}

We firstly estimated the fluorescent X-ray flux due to the scattering of solar X-rays by
the upper Martian atmosphere. This component depends on the amount of the incident solar 
X-rays during the observation. 
Figure \ref{fig:solar-xray} shows the GOES-12 solar X-ray flux in 2001-2008. 
The Suzaku observation corresponds to the solar minimum phase when
the solar X-ray flux is significantly lower than those in the other two Chandra and 
XMM-Newton observations. Close-up views during the three observations are shown in 
figure \ref{fig:solar-xray-CNO}. We averaged the solar X-ray fluxes during the observation
time. The average solar X-ray fluxes were 3.1 $\times$ 10$^{-7}$ W m$^{-2}$ (Chandra), 
2.5 $\times$ 10$^{-6}$ W m$^{-2}$ (XMM-Newton) and 3.1 $\times$ 10$^{-8}$ W m$^{-2}$ (Suzaku). 
Thus, the solar X-ray at the Suzaku observation is 10 times lower than the Chandra and 
100 times lower than the XMM-Newton. 

\begin{figure}[p]
\begin{center}
\FigureFile(80mm,80mm){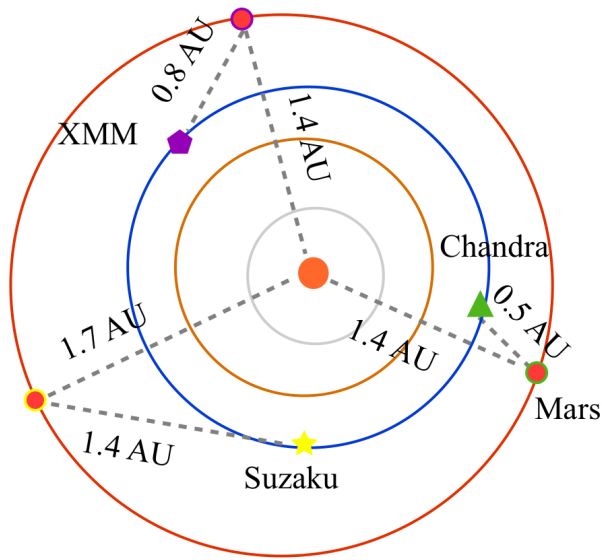}
\end{center}
\caption{Positions of Sun, Mars and X-ray observatories in the three observations 
(http://space.jpl.nasa.gov/). 
An orange circle indicates Sun, while red circles represent Mars. 
A blue circle shows an orbit of Earth, and symbols along the blue circle shows 
positions of the three observatories. A pentagon is XMM-Newton, 
a triangle is Chandra and a star is Suzaku.}
\label{fig:orbit}
\end{figure}

The observed fluorescent X-ray flux from Mars in O\emissiontype{VII} band 
(0.5--0.65 keV) derived from the section 3.2 of \citet{den02a} and the column (c) in table 2 
of \citet{den06} with the Suzaku response was 5.5$\times$10$^{-5}$ and 
3.7$\times$10$^{-5}$ cts cm$^{-2}$ s$^{-1}$ in the Chandra and XMM-Newton observations, 
respectively. 
By taking into account the solar X-ray flux, the phase angle, and the distance of Earth-Mars 
and Sun-Mars, the fluorescent X-ray flux during the Suzaku observation is estimated.
The positions of the X-ray observatories, Sun and Mars in the three observations 
are shown in figure \ref{fig:orbit}. The phase angle was 18.2, 41.2 and 37.0 degree in 
the Chandra, XMM-Newton, and Suzaku observations, respectively, 
while the geocentric distance of Mars was 0.5, 0.8 and 1.4 AU, and the heliocentric distance 
was 1.4, 1.4, and 1.7 AU.
Provided that the phase angle is the same as that in the Suzaku observation from 
figure 12 in \citet{den02a}, the photon fluxes of the Chandra and XMM-Newton observations,
are converted into 4.9$\times$10$^{-5}$ and 3.8$\times$10$^{-5}$ cts cm$^{-2}$ s$^{-1}$, respectively.
In addition, by considering the solar X-ray flux, and the geocentric and heliocentric distances, 
the expected photon fluxes become $4.2\times10^{-7}$ and $1.1\times10^{-7}$ cts cm$^{-2}$ s$^{-1}$, 
respectively. 
These values are more than an order of magnitude lower than the 2$\sigma$ upper limit 
of 4.3$\times$10$^{-5}$ ph cm$^{-2}$ s$^{-1}$. Therefore, the scattering of solar 
X-rays during the Suzaku observation should be small. 

\subsection{Solar wind charge exchange}
\label{sec:sec:swcx}

We next estimate the charge exchange emission.
Here, we simply assume that only a single charge exchange 
occurs while the highly charged ions in the solar wind go through 
the Martian atmosphere, from the analogy with the charge exchange
emission of moderately active comets and the Earth's exosphere \citep{cra97,cra01}.
We can express the 
X-ray volume emissivity $P_{\rm x} = \alpha n_{\rm sw} u_{\rm sw} n_{\rm n}$ [eV cm$^{-3}$ s$^{-1}$],  
where $n_{\rm sw}$, $u_{\rm sw}$, and $n_{\rm n}$ are the solar wind 
proton density, solar wind speed, and neutral target density that is in 
the collisionally thin regions of the Martian exosphere, respectively. 
The parameter $\alpha$ is given as 
$\alpha \approx f_{\rm h} <\sigma_{\rm cx}> E_{\rm ave}$, where 
$f_{\rm h}$ is a fraction of heavy ions in the solar wind, $<\sigma_{\rm cx}>$
is an average cross section of charge exchange for all species and charge 
state, and $E_{\rm ave}$ is an average photon energy.
By integrating this equation in the line of sight and dividing it by the solid angle, 
we can obtain the X-ray energy flux.
Hence, we assume that the X-ray flux is proportional to the incident solar wind flux
and the exospheric density integrated in the line of sight.
According to a Chamberlain exospheric model \citep{cham87},  
the exospheric density will depend on the density and temperature at the exobase. 
Here we assume the fixed scale hight for simplicity. Thus, when the density is 
increased by inserting additional particles into the atmosphere, the charge exchange 
X-ray flux would also increase.
Therefore, the X-ray observations can provide a means of remotely estimating 
the condition of the exosphere.

\begin{figure}[p]
\begin{center}
\FigureFile(120mm,100mm){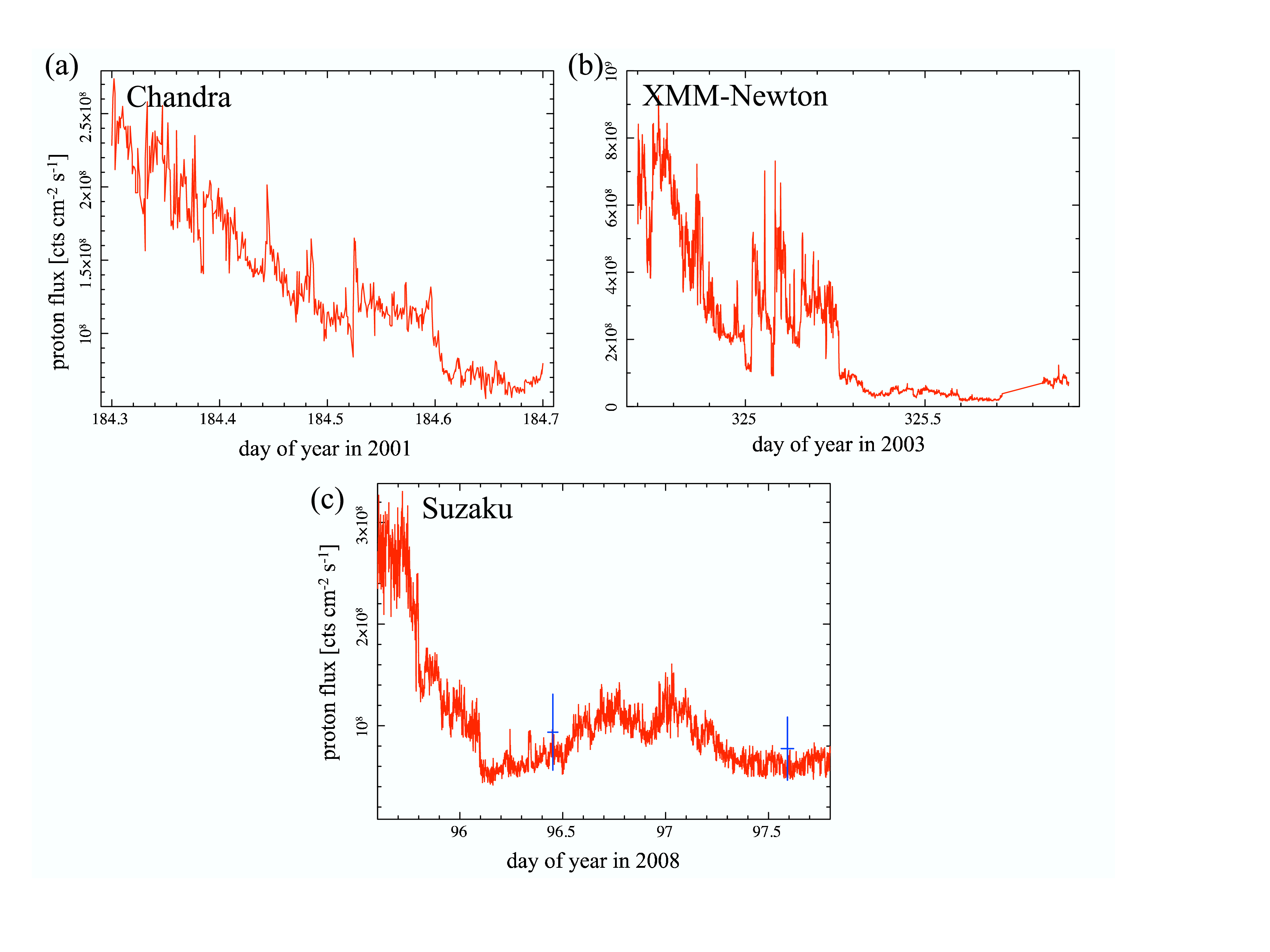}
\end{center}
\caption{The solar wind proton flux measured by the ACE satellite 
corresponding to the Chandra (panel a), XMM-Newton (panel b), and Suzaku (panel c) observations. The observed data is shifted in time considering the propagation time 
of the solar wind to Mars.  In situ solar wind flux observed by ASPERA-3 is shown 
by blue symbols.}
\label{fig:sum-pro}
\end{figure}

Below we place an upper limit on the exospheric density by comparing the Suzaku
observation with the past Chandra and XMM-Newton observations.
We used the solar wind proton flux obtained with ACE SWEPAM\footnote{http://www.srl.caltech.edu/ACE/ASC/level2/index.html} to estimate the incident solar wind heavy ions.
Although the proton flux, which is a proxy for the heavy ions in the solar wind,
is not very strongly correlated with the heavy ion flux \citep{neu00}, 
the sparse and not always available ACE SWICS O$^{+7}$ data during the X-ray observations
hindered us to directly estimate the heavy ion flux.
In addition to these data, we used the in situ solar wind proton flux 
using the ASPERA-3 instrument onboard Mars Express \citep{bara06}. While the data on 
April 3 was lost because of the overwrite in the onboard data recorder, we get the solar wind flux 
on April 4 and 5 (blue symbols in figure \ref{fig:sum-pro}c). The fluxes are consistent with the estimated fluxes using ACE.
Note that Mars Express was not in orbit during the Chandra and XMM-Newton 
observations. 
We considered arrival times of the solar wind depending on the locations 
of Earth, Mars and the ACE satellite that orbits around a Lagrangian point (L1) 
between the Sun and Earth (1.5 million km away from Earth), when we estimate 
the average proton flux. 
When Mars is in the upstream side of the solar wind compared to Earth,
the solar wind which hits Mars will arrive Earth at $t_{\rm E}=t_{\rm M}+\Delta t$, 
where $t_{\rm E}$, $t_{\rm M}$ and $\Delta t$ are the arrival times at Earth and Mars, 
and the expected time delay, respectively.
The expected time delay can be expressed as 
$\Delta t = \theta/\Omega - (\Delta r_{\rm EM} + \Delta r_{\rm EA})/u_{\rm sw} $ based on 
 \cite{neu00}, where $\theta$ is the difference in the heliographic longitudes of Earth  and Mars, 
$\Omega = 14.7^{\circ}$/day is the rotational speed of the Sun, 
$\Delta r_{\rm EM}$ is the difference of the heliocentric distances of Earth and Mars, and 
$\Delta r_{\rm EA} = 1.5\times10^6$ [km] is the distance between Earth and the ACE satellite.
We thus estimated the proton density and speed at the time of the Mars observations
considering the time delay. Moreover, we corrected the proton density for the difference
of the heliocentric distances of Mars and the ACE satellite.
Figure \ref{fig:sum-pro} shows the solar wind proton flux at $t_{\rm M}$ in
the Chandra, XMM-Newton and the Suzaku observations.
Table \ref{tab:sum-obs} summarizes the average proton flux at $t_{\rm M}$, 
distances, and the observed X-ray fluxes with the three observatories. 

We then compared the 0.5--0.65 keV X-ray flux during the Suzaku observation 
with those in the Chandra and XMM-Newton observations. 
The 0.5--0.65 keV X-ray fluxes in the Chandra and XMM-Newton observations 
were estimated from the halo model in section 3.2 of \citet{den02a} 
and in the column (c) in table 2 of \citet{den06}, respectively.
Simply from the geocentric distances, the expected X-ray flux during the Suzaku observation 
becomes 0.1 and 0.3 times smaller than those of Chandra and XMM-Newton, respectively. 
The proton flux corresponding to the Suzaku observation was 0.8 and 0.5 times 
those of Chandra and XMM-Newton. Then, the expected X-ray flux during the 
Suzaku observation is still 0.1 and 0.2 times smaller than those of Chandra 
and XMM-Newton, respectively. 
Therefore, if we assume that the solar wind heavy ion flux scaled with the proton flux 
and that the ionization states were similar during the three observations, 
the undetection of the Martian halo emission with Suzaku means that the exospheric 
density at solar minimum is not significantly higher than those at solar maximum. 
We estimate an upper limit of the exospheric neutral density during the Suzaku 
observation as 70 and 6 times during the Chandra and XMM-Newton observations.
This result places a 2$\sigma$ limit on the modeling of the Martian exosphere 
and can not reject the higher exospheric density at lower solar activity. 

\section{Summary}
\label{sec:summary}

We investigated the Martian X-ray emission with Suzaku XIS in 2008 April. 
For the first time, we observed Mars at solar minimum. We carefully excluded possible
point sources from the XIS 0.2--1 and 1--5 keV images and corrected the data for the point 
sources and the orbital motion of Mars. We analyzed the X-ray image and 
spectrum, and concluded that there was no significant X-ray flux from Mars 
observed by Suzaku. 
We placed stringent upper limits on the Martian X-ray flux around 
the O\emissiontype{VII} band (0.5--0.65 keV). 
We compared the upper limit with the past Chandra and XMM-Newton detections of the Martian
X-ray emission consisting of the scattering of solar X-rays by the planet's body and the 
solar wind charge exchange halo. 
Using the simultaneously obtained solar X-ray and wind data, we estimated that the fluorescent 
emission due to the scattering must be negligible, and placed an upper limit on the Martian 
exospheric density from the X-ray data as 70 and 6 times during the Chandra
and XMM-Newton observations, respectively. \\
~~~~~~{\bf{Acknowledgments.}}
We thank the ACE SWEPAM instrument team and Dr. K. Dennerl for valuable comments. 
This work was supported by Grant-in-Aid for JSPS Fellows.

\clearpage


\begin{table}[p]
\begin{center}
\caption{Summary of the time of the individual pointings.}
\begin{tabular}{cccc}\hline\hline
Pointing ID & Begin [UT]             & End [UT]                 & Exposure time [s] \\ \hline
1                  & April 3 ~08:11:44 & April 3 ~09:50:19 & 2913 \\
2                  & April 3 ~09:50:32 & April 3 ~11:50:19 & 4887 \\                               
3                  & April 3 ~11:50:28 & April 3 ~13:50:19 & 3910 \\                               
4                  & April 3 ~13:50:32 & April 3 ~15:50:19 & 2288 \\                               
5                  & April 3 ~15:50:28 & April 3 ~17:50:16 & 1368 \\                               
6                  & April 3 ~17:50:33 & April 3 ~19:50:18 & 1939 \\                               
7                  & April 3 ~19:50:31 & April 3 ~21:40:18 & 2511 \\                               
8                  & April 3 ~21:40:31 & April 4 ~00:10:24 & 5608 \\                               
9                  & April 4 ~00:10:37 & April 4 ~01:50:24 & 2175 \\                               
10                & April 4 ~01:50:33 & April 4 ~03:50:24 & 3772 \\                               
11                & April 4 ~03:50:33 & April 4 ~05:50:24 & 4620 \\                               
12                & April 4 ~05:50:37 & April 4 ~07:50:24 & 3455 \\                               
13                & April 4 ~07:50:33 & April 4 ~09:50:24 & 4252 \\                               
14                & April 4 ~09:50:37 & April 4 ~11:50:24 & 4825 \\                               
15                & April 4 ~11:50:33 & April 4 ~13:50:24 & 3831 \\                               
16                & April 4 ~13:50:37 & April 4 ~15:50:24 & 1440 \\                               
17                & April 4 ~15:50:33 & April 4 ~17:50:20 & 1978 \\                               
18                & April 4 ~17:50:37 & April 4 ~19:50:04 & 1862 \\                               
19                & April 4 ~19:50:21 & April 4 ~21:50:14 & 1700 \\                               
20                & April 4 ~21:50:27 & April 4 ~23:50:14 & 1849 \\                               
21                & April 4 ~23:50:23 & April 5 ~01:50:14 & 2881 \\                               
22                & April 5 ~01:50:27 & April 5 ~03:50:14 & 4025 \\                               
23                & April 5 ~03:50:27 & April 5 ~05:50:14 & 4519 \\                               
24                & April 5 ~05:50:27 & April 5 ~07:50:14 & 3423 \\                               
25                & April 5 ~07:50:23 & April 5 ~09:50:14 & 4261 \\                               
26                & April 5 ~09:50:27 & April 5 ~11:50:14 & 4798 \\                               
27                & April 5 ~11:50:23 & April 5 ~13:00:13 & 1100 \\
\hline
\end{tabular}
\label{tbl:obs}
\end{center}
\end{table}

\begin{table}[p]
\begin{center}
\caption{2$\sigma$ upper limits of the Martian X-ray emission$^a$.}
\begin{tabular}{ccc}\hline\hline
energy band [keV]   & count rate [cts s$^{-1}$]  & photon flux [photon cm$^{-2}$ s$^{-1}$] \\
\hline
0.5--0.65           & 1.7$\times10^{-3}$      & 4.3$\times10^{-5}$ \\
0.2--1.0            & 4.3$\times10^{-3}$      & 2.3$\times10^{-4}$ \\
1.0--5.0            & 6.7$\times10^{-3}$      & 3.5$\times10^{-5}$ \\
\hline
\end{tabular}
\bigskip

{
$^a$ The BI data was used for 0.5--0.65 and 0.2--1.0 keV, while the FI was used for 1.0--5.0 keV.
}
\label{tbl:up}
\end{center}
\end{table}

\begin{table}[p]
\begin{center}
\caption{Summary of solar activity for Chandra, XMM-Newton and Suzaku. X-ray flux is in 0.5-0.65 keV.}
\begin{tabular}{ccccc}\hline\hline
            & observation     & proton flux$^{a}$                     & Earth-Mars/Sun-Mars & X-ray flux \\
            &   date                & (cm$^{-2}$ s$^{-1}$)   & (AU)                                & (photon cm$^{-2}$ s$^{-1}$)\\\hline
Chandra     & 2001/07/04      & 1.3$\times10^8$        & 0.5/1.4             & 5.8$\times10^{-6}$  \\
XMM-Newton  & 2003/11/19-21   & 2.2$\times10^8$        & 0.8/1.4             & 3.8$\times10^{-5}$  \\
Suzaku      & 2008/04/03-05   & 1.0$\times10^8$        & 1.4/1.7             & $<$4.3$\times10^{-5}$ \\\hline
\end{tabular}
\bigskip

{
$^a$ The proton flux is estimated by considering the arrival time and the heliospheric distance (see text).
}
\label{tab:sum-obs}
\end{center}
\end{table}

\clearpage


\clearpage

\begin{thebibliography}{}

\bibitem[Anderson(1974)]{and74}
  Anderson, D. E. 1974, J. Geophys. Res., 79, 1513.
\bibitem[Barabash et al.(2006)]{bara06}
  Barabash, S., et al. 2006, Space Science Reviews, 126, 113.
\bibitem[Chamberlain \& Hunten(1987)]{cham87}
  Chamberlain, J.W., Hunten, D.M. 1987, Theory of Planetary Atmospheres, 
  (San Diego, CA: second ed Academic Press), 335
\bibitem[Chaufray et al.(2008)]{chau08}
  Chaufray, J.Y., Bertaux, J.L., Leblanc, F., Qu\'{e}merais, E. 2008, Icarus, 195, 598.
\bibitem[Cravens(1997)]{cra97}
  Cravens, T.E. 1997, J. Geophys. Res. Lett., 24, 105.
\bibitem[Cravens et al.(2001)]{cra01}
  Cravens, T.E., Robertson, I.P., \& Snowden, S.L., 2001, J. Geophys. Res. 106, 24883.
\bibitem[Dennerl et al.(1997)]{den97}
  Dennerl, K., Englhauser, J., \& Tr\"{u}mper, J. 1997, Science, 277, 1625.
\bibitem[Dennerl(2002)]{den02a}
  Dennerl, K. 2002, A\&A, 394, 1119.
\bibitem[Dennerl et al.(2002)]{den02b}
  Dennerl, K., Burwitz, V., Englhauser, J., Lisse, C., \& Wolk, S. 2002, A\&A, 386, 319.
\bibitem[Dennerl et al.(2006)]{den06}
  Dennerl, K., et al. 2006, A\&A, 451, 709.
\bibitem[Ezoe et al.(2010)]{ezo10}
  Ezoe, Y., Ishikawa, K., Ohashi, T., Miyoshi, Y., Terada, N., 
  Uchiyama, \& Y., Negoro, H. 2010, ApJ, 709, L178.
\bibitem[Galli et al.(2006)]{gal06}
  Galli, A., et al. 2006, Space Science Reviews, 126, 447.
\bibitem[Giacconi et al.(2001)]{gia01}
  Giacconi, R., et al. 2001, ApJ, 551, 624.
\bibitem[Gladstone et al.(2002)]{gla02}
  Gladstone, G.R., et al. 2002, Nature, 415, 1000.
\bibitem[Grader et al.(1968)]{gra68}
  Grader, R.J., Hill, R.W., \& Seward, F.D. 1968, J. Geophys. Res. 73, 7149.
\bibitem[Holmstr\"{o}m and Barabash(2001)]{hol01}
  Holmstr\"{o}m, M., \& Barabash, S. 2001, Geophys. Res. Lett., 28, 1287.
\bibitem[Koyama et al.(2007)]{koy07}
  Koyama, K., et al. 2007, PASJ, 59, S23.
\bibitem[Krasnopolsky \& Gladstone(1996)]{kra96}
  Krasnopolsky, V.A., \& Gladstone, G.R. 1996, J. Geophys. Res., 101, 765.
\bibitem[Krasnopolsky(2002)]{kra02}
  Krasnopolsky, V.A. 2002, J. Geophys. Res., 107, 5128.
\bibitem[Krasnopolsky et al.(2004)]{kra04}
  Krasnopolsky, V.A., Greenwood, J.B., \& Stancil, P.C. 2004, Space Sci. Rev., 113, 271.
\bibitem[Mitsuda et al.(2007)]{mit07}
  Mitsuda, K., et al. 2007, PASJ, 59, S1. 
\bibitem[Ness et al.(2004)]{nes04}
  Ness, J.-U., Schmitt, J.H.M.M., \& Robrade, J. 2004, A\&A, 414, L49.
\bibitem[Neugebauer et al.(2000)]{neu00}
  Neugebauer, T., Cravens, T.E., Lisse, C.M., Ipavich, F.M., Christian, D.,
  von Steiger, R., Bochsler, P., Shah, P.D., \& Armstrong, T.P. 2000, J. Geophys. Res., 105, 20,949.
\bibitem[Serlemitsos et al.(2007)]{ser07}
  Serlemitsos, P.J., et al. 2007, PASJ, 59, S9.  
\bibitem[Uchiyama et al. 2008]{uch08}
  Uchiyama, Y., et al. 2008, PASJ, 60, S35. 
\bibitem[Wegmann et al.(1998)]{weg98}
  Wegmann, R., Schmidt, H.U. , Lisse, C M., Dennerl, K., \& Englhauser, J. 1998, Planet. Space. Sci., 46, 603

\end{thebibliography}
\end{document}